\documentclass[conference]{IEEEtran}

\usepackage[nolist]{acronym}
\usepackage{subcaption}
\usepackage{tikz}
\usetikzlibrary{shapes, arrows, calc, positioning}
\usepackage{pgfplots}
\usepackage{pgfplotstable}
\usepgfplotslibrary{fillbetween}
\pgfplotsset{compat=1.18}
\usepgfplotslibrary{groupplots}
\usepackage{tabularx}
\usepackage{cite}
\usepackage{url}
\usepackage{booktabs}
\usepackage[cmex10]{amsmath}
\usepackage[final]{changes}
\usepackage{caption}
\usepackage{eurosym}
\usepackage{tabularx}
\usepackage{subcaption}\usepackage{amssymb}

\definecolor{darkblue}{HTML}{332288}
\definecolor{green}{HTML}{117733}
\definecolor{turq}{HTML}{44AA99}
\definecolor{lightblue}{HTML}{88CCEE}
\definecolor{sand}{HTML}{DDCC77}
\definecolor{pink}{HTML}{CC6677}
\definecolor{purple}{HTML}{AA4499}
\definecolor{darkpurple}{HTML}{882255}

\begin{document}

\title{Optimal Battery Bidding under Decision-Dependent State-of-Charge Uncertainties}


\author{
\IEEEauthorblockN{Jan Brändle\IEEEauthorrefmark{1}\IEEEauthorrefmark{2}, Gabriela Hug\IEEEauthorrefmark{1}}
\IEEEauthorblockA{\IEEEauthorrefmark{1}Power Systems Laboratory, ETH Zürich, Zürich, Switzerland}
\IEEEauthorblockA{\IEEEauthorrefmark{2}janbr@ethz.ch}
\vspace{-0.4cm}
}

\begin{acronym}
  \acro{BESS}{Battery Energy Storage System}
  \acro{BMS}{Battery Management System}
  \acro{DAM}{Day-Ahead Market}
  \acro{FCR}{Frequency Containment Reserves}
  \acro{LFP}{Lithium Iron Phosphate}
  \acro{MPC}{Model Predictive Control}
  \acro{OCV}{Open-Circuit Voltage}
  \acro{SOC}{State of Charge}
\end{acronym}

\maketitle

\pagestyle{plain}

\begin{abstract}
Lithium Iron Phosphate (LFP) Battery Energy Storage Systems (BESSs) are a key enabler of the energy transition. However, they are known to exhibit significant inaccuracies in the estimation of their State of Charge (SOC). Such estimation errors can directly impact the participation of BESSs in electricity markets. In this work, we demonstrate that neglecting SOC uncertainty in battery bidding can lead to significant delivery failures, including the inability to meet promised frequency reserves. To address this risk, we investigate bidding strategies that account for SOC uncertainty. We propose three constraint-tightening optimization approaches of increasing complexity: (i) a fixed-margin formulation, (ii) an adaptive-margin optimizer, and (iii) an uncertainty-aware optimization model. The latter explicitly accounts for the decision-dependent nature of the uncertainty. Numerical results demonstrate that while all three approaches robustify against SOC uncertainty, the uncertainty-aware formulation outperforms the others in maximizing revenue while ensuring reliable frequency reserve provision. This highlights the significance of treating SOC uncertainty as an endogenous process within the operational strategy.
\end{abstract}

\begin{IEEEkeywords}
Ancillary services, Battery energy storage systems, Bidding strategies, Decision-dependent uncertainty, State-of-charge estimation.
\end{IEEEkeywords}

\section{Introduction}\label{sec:introduction}
Energy storage systems are a key enabler for the integration of variable renewable energy sources and are essential to meet the increased flexibility requirements of future energy systems~\cite{Koolen2023}. Among these, lithium-ion batteries have emerged as the predominant technology for new installations \cite{GonzalezCuenca2025}. Specifically, due to their low cost, extended cycle life, and enhanced safety, \ac{LFP} batteries have become the dominant chemistry for grid-scale applications, representing approximately 80\% of the market in 2023 for storage applications \cite{IEA2024}.

It is well established that grid-scale \acp{BESS} are effective for providing ancillary services, such as \ac{FCR} \cite{Oudalov2007}. Indeed, FCR typically provides a primary revenue stream for batteries participating in electricity markets \cite{IRENA2019, Figgener2020}. Recently, beyond FCR, participation in secondary frequency reserves and multi-use operation, combining reserves with spot market arbitrage, have emerged as profitable strategies \cite{Engl2020}. 

For such bidding processes, accurate \ac{SOC} estimation is essential. However, for \ac{LFP} systems, this estimation remains challenging due to the characteristic voltage plateau and significant hysteresis of their \ac{OCV} curves \cite{Joest2024}. In particular, continued participation in ancillary services, especially \ac{FCR}, can aggravate \ac{SOC} estimation errors \cite{Joest2024, Powin2025}. This is due to the operational requirements of \ac{FCR}, where \acp{BESS} maintain a mid-range \ac{SOC} to provide bidirectional availability, rarely leaving the 20\% to 80\% \ac{SOC} range \cite{Jacque2022, Thien2017}. By operating away from the high-curvature regions (or "knees") of the voltage curve at high or low \ac{SOC}, this application lacks the voltage gradients that allow typical estimation algorithms to recalibrate their state estimates. As a consequence, \ac{SOC} errors can accumulate and reach or even exceed 20\% of the battery capacity \cite{Accure2024}. Such inaccuracies in \ac{SOC} can have severe implications, as erroneous state estimates lead to suboptimal dispatch decisions and to financial penalties for failing to comply with contractual obligations \cite{Powin2025}. In the most extreme cases, should a \ac{BESS} not be able to deliver the contractually agreed services to the grid, it could lose prequalifications and market access \cite{EC2017}.

Addressing \ac{SOC} estimation drift in \ac{LFP} batteries, particularly in applications such as \ac{FCR}, remains an ongoing research challenge. Current literature and industry address this problem through two avenues: First, there is a push to improve the estimation process by exploring more advanced \ac{SOC} estimation algorithms that go beyond standard Coulomb counting and \ac{OCV}-based methods. These algorithms can be physics-based \cite{Schwunk2013, Li2013, Li2020} or data-driven estimation techniques~\cite{Hoss2020, How2020}. However, they still rely on current integration to some extent and cannot reliably eliminate estimation drift \cite{Accure2024}. Second, cloud-based predictive analytics use cloud computing to support higher model fidelity and fleet-wide comparisons~\cite{Accure2024, Twaice2024}. Yet, such solutions require an extensive sensing and data acquisition ecosystem and a representative battery population to be effective. 

Although these approaches may provide improvements in the precision of estimation, their focus is solely on estimation and data analytics. In contrast, the potential of operational strategy as a mechanism to improve SOC accuracy has received comparatively little attention. Since the \ac{SOC} error can be recalibrated by targeting specific SOC operating points, we can treat the estimation uncertainty as decision-dependent and integrate it into the bidding strategy.

Thus, both the treatment of \ac{SOC} uncertainty in battery bidding and the potential for proactive bidding strategies that manage this uncertainty remain underinvestigated. To address these research gaps, we investigate a robust treatment of \ac{SOC} uncertainty in battery bidding and introduce a novel, uncertainty-aware optimization that actively selects actions to reduce uncertainty. Hence, the contributions of this paper are threefold:
\begin{itemize} \item We formulate three constraint-tightening approaches for bidding under SOC uncertainty and provide a comparative analysis of their performance.
\item We analyze the trade-off between revenue and robustness, and highlight the impact of SOC uncertainty on FCR compliance. 
\item We show that treating SOC error as decision-dependent allows the operational strategy of the battery to actively reduce uncertainty and to significantly improve the balance between revenue and compliance.
\end{itemize}

\added{These contributions are particularly relevant for battery operators participating in reserve markets, though the framework extends to any application where strategic scheduling of SOC recalibration events is beneficial.}

The remainder of this paper is structured as follows: In Section \ref{sec:methodology}, we first detail the \ac{SOC} error dynamics, explain the proposed \ac{MPC} optimization framework, and introduce three constraint-tightening formulations of battery bidding. In Section \ref{sec:caseStudy}, we introduce the case study focused on the German \ac{DAM} and \ac{FCR} market. In Section \ref{sec:results}, we compare the performance of the proposed optimizers. Finally, in Section \ref{sec:conclusion}, we summarize the most important conclusions.

\section{Methodology}\label{sec:methodology}
In this section, we detail the battery model and \ac{SOC} error dynamics. We then develop the battery and market models for the optimization framework and propose three constraint-tightening formulations designed to account for \ac{SOC} uncertainty.

\subsection{SOC Error Model and \ac{BESS} Simulator}
The \ac{SOC} of a \ac{BESS} is not a directly observable quantity, but rather needs to be continuously estimated by the \ac{BMS} from current and voltage measurements. Since this estimation is error-prone, we need to distinguish between the true physical \ac{SOC}, $s^{\text{true}} \in [0,1]$, and the value reported by the \ac{BMS}, $s^{\text{rep}}$. Their relationship is given by
\begin{equation}
    s^{\text{true}}_t = s^{\text{rep}}_t + w_t.
\end{equation}



When planning battery operation in the optimal bidding problem, we only have access to the reported value. As established in \cite{Powin2025}, the estimation error $w_t$ can be modeled as
\begin{align}
    w_{t+1} &= a(s_t^{\text{true}}) \cdot \Big( w_t + |p^{true}_t| \cdot \eta \Big),
\end{align}
where $p_t^{\text{true}}$ denotes the power dispatched at time $t$, and $\eta \sim \mathcal{N}(\beta, \sigma^2)$ is a stochastic term that represents measurement bias and variability. The function $a$ is piecewise linear in the true \ac{SOC} and is given by
\vspace{-0.3cm}
\begin{figure}[h!]
  \centering
  \begin{tabular}{cc}
    \begin{minipage}{0.5\columnwidth}
    \vspace{-0.3cm}
    \scalebox{0.95}{
        $a(s) =
    \begin{cases}
    \dfrac{s}{b}, & s \le b, \\
    1, & b < s \le c, \\
    1 - \dfrac{s - c}{1-c}, & s > c.
    \end{cases}$
    }
    \end{minipage} 
    & 
    \begin{minipage}{0.4\columnwidth}
        \includegraphics[width=\linewidth]{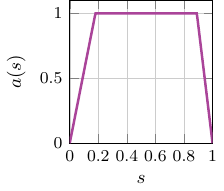}
    \end{minipage}
  \end{tabular}
\end{figure}
\vspace{-0.1cm}

This error model reflects the inherent characteristics of \ac{SOC} estimation in \ac{LFP} batteries: Whenever the \ac{SOC} lies in a range between a lower threshold $b$ (typically around 15\% to 20\%) and an upper threshold $c$ (of around 90\%), the voltage barely changes and thus provides only limited information about the \ac{SOC} \cite{Joest2024}. As a consequence, the error accumulates. In contrast, the rapidly decreasing scaling factor towards the extremes of the \ac{SOC} region reflects that the estimation can be recalibrated and more certainty can be gained in high/low \ac{SOC} regions. For the scope of this project, we additionally assume that there is an upper bound on the error $w_t$, i.e., that the error saturates at a certain level and does not increase indefinitely.

Complementing the error model, we use a battery model to simulate the true \ac{SOC} trajectory as a function of the realized charging and discharging powers. To maintain physical consistency, we distinguish between the commanded power dispatch $p_t$ (representing the output of the optimization routine) and the realized power dispatch $p^{\text{true}}_t$, decomposed into discharging and charging variables as $p^{\text{true}}_t = p^{\text{true}}_{\text{dis},t} - p^{\text{true}}_{\text{ch},t}$. The latter represents the actual power output, which is limited by the true available energy and prevents \ac{SOC} violations beyond the physical limits. For a time step $t$ of duration $\Delta t$, the state transition is modeled as:
\begin{align}
    s^{\text{true}}_{t+1} &= s^{\text{true}}_t + \frac{\Delta t}{C} \left( \eta_\text{ch} \, p^{\text{true}}_{\text{ch},t} - \eta_\text{dis}^{-1} \, p^{\text{true}}_{\text{dis},t} \right), \label{eq:soc_evolution} \\
    s^{\text{true}}_{0} &= s_{\text{init}}, \label{eq:soc_initial}
\end{align}
where $\eta_\text{ch}$ and $\eta_\text{dis}$ denote the charging and discharging efficiencies, respectively, and $C$ is the energy capacity of the battery system.

\subsection{Market and \ac{MPC} Framework}
\begin{figure}[b]
    \centering
    \vspace{-0.3cm}
    \includegraphics[width=1\linewidth]{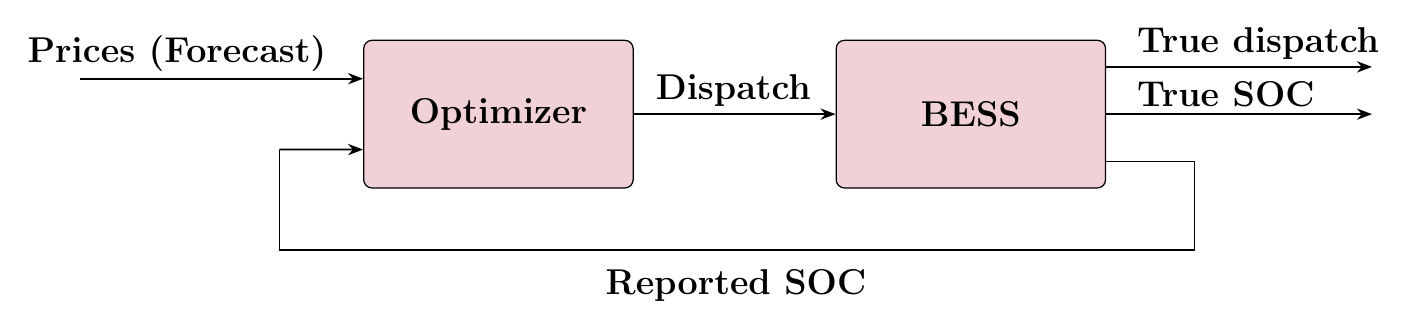}
    \caption{Schematic of the optimization framework: The optimizer takes hourly decisions about dispatch and bids. The BESS simulates the dispatch and the SOC error dynamics, and provides a reported SOC estimate to the optimization.}
    \label{fig:diagram}
\end{figure}

We formulate the bidding process as a receding horizon optimization problem, which adjusts dispatch decisions hourly and takes bidding decisions at each market auction instance, as illustrated in Fig.~\ref{fig:diagram}. This section details the battery model, market constraints and cost function that constitute the \ac{MPC} optimization framework.
\subsubsection{Battery model for the optimization}
The optimizer's internal battery model follows the same dynamics as described in~\eqref{eq:soc_evolution} and \eqref{eq:soc_initial}. Here, the \ac{SOC} and dispatch levels serve as decision variables, with the optimization initialized using the reported \ac{SOC} value. To maintain physical consistency, we ensure mutual exclusivity of charging and discharging with binary constraints and constrain the power within the battery's rated capacity limits.

\subsubsection{Market Framework and bidding constraints}
For the purpose of this paper, we model the simultaneous participation of the \ac{BESS} in two distinct markets: the day-ahead spot market and the primary frequency reserve market. In the first, the battery can buy or sell energy. It optimizes its market bids for time instances $t$, denoted by $p^{\text{DA}}_{\text{bid},t}$. Assuming that these bids are accepted, any deviation between the scheduled and the realized power dispatch results in an imbalance, $p^{\text{imb}}_{t}$, which can be split into positive and negative imbalances. The sum $p_t$ of these powers constitutes the power provided or absorbed by the battery, i.e.,
\begin{align}
p_{t} &= p^{\text{DA}}_{\text{bid},t} + p^{\text{imb}}_{t}, \\
p^{\text{imb}}_{t} &= p^{\text{imb}}_{\text{pos},t} - p^{\text{imb}}_{\text{neg},t}.
\end{align}
Imbalances are generally discouraged and, within the scope of this paper, imbalance prices are always assumed to be unfavorable compared to the corresponding spot market prices. 

Secondly, the battery can sell reserve generation capacity in reserve markets. In this work, the European \ac{FCR} market is considered. For time instances $t$, the battery optimizes its bids into this market, $p^{\text{FCR}}_{\text{bid},t}$. Assuming that these bids are accepted, the battery needs to ensure that it can fulfill its \ac{FCR} commitments without violating physical power or energy limits. Therefore, the following set of constraints is imposed:
\begin{align}
    p_{\text{ch},t} &\le P_{\max} - p^{\text{FCR}}_t, \label{eq:fcr_p_ch} \\
    p_{\text{dis},t} &\le P_{\max} - p^{\text{FCR}}_t, \label{eq:fcr_p_dis} \\
    s_t, s_{t+1} &\le s_{\max} - \frac{\eta_{\text{ch}}\Delta t_{\text{FCR}}}{C} p^{\text{FCR}}_t := \tilde{s}_{\max,t}, \label{eq:fcr_s_max} \\
    s_t, s_{t+1} &\ge s_{\min} + \frac{\Delta t_{\text{FCR}}}{\eta_{\text{dis}} C} p^{\text{FCR}}_t := \tilde{s}_{\min,t}. \label{eq:fcr_s_min}
\end{align}
The first two constraints ensure that sufficient power capacity is available to enable full FCR activation, while accommodating the charging and discharging required by the spot market. The last two constraints define the effective \ac{SOC} limits, $\tilde{s}_{\max,t}$ and $\tilde{s}_{\min,t}$, which ensure that enough energy is available for a full \ac{FCR} activation over a maximum duration of $\Delta t_{\text{FCR}}$. These buffers are enforced at both the beginning and end of the interval $[t, t+1]$ to guarantee enough margin for activation at any instance during the interval. Any deviation between the submitted FCR bid, $p^{\text{FCR}}_{\text{bid},t}$, and the delivered FCR, $p^{\text{FCR}}_{t}$, is denoted by the shortfall variable $p^{\text{FCR}}_{\text{short},t}$.

Beyond energy buffers, the optimization must respect the temporal structure of the different market products. In the case of \ac{FCR}, bids must be submitted in four-hour blocks, and are therefore constrained as such. Since the optimization operates with a receding horizon, certain market bids may have been fixed in previous time steps already. These variables are treated as fixed parameters and are constrained to the power values awarded in prior gate closures.

\subsubsection{Optimization objective}
At each time step, the optimizer maximizes the net profit: 
\begin{equation}
\begin{aligned}
J = \sum_{t=0}^{T-1} & \bigg( \pi^{\text{DA}}_t p^{\text{DA}}_{\text{bid},t} + \pi^{\text{FCR}}_t p^{\text{FCR}}_{\text{bid},t} - c_{\text{deg}} \left( p_{\text{dis},t} + \zeta p^{\text{FCR}}_{t} \right)\\
& - c_{\text{imb}} (p^{\text{imb}}_{\text{pos},t} + p^{\text{imb}}_{\text{neg},t}) - c_{\text{fcr}} p^{\text{FCR}}_{\text{short},t} \bigg),
\end{aligned} \label{eq:objective_function}
\end{equation}
where $T$ is the optimization horizon and $\pi^{\text{FCR}}_t$ and $\pi^{\text{DA}}_t$ denote the \ac{FCR} and \ac{DAM} prices, respectively. The first two terms represent the revenue from \ac{FCR} capacity payments and \ac{DAM} bidding, assuming that the bids are accepted. The third term approximates the cost of battery degradation, where $\zeta = 0.1$ denotes the average \ac{FCR} activation. The last two terms impose penalties on imbalances on the \ac{DAM} and shortfalls in \ac{FCR} via the cost factors $c_{\text{imb}}$ and $c_{\text{fcr}}$.

\subsection{Constraint-Tightening \ac{MPC} formulation}
Given perfect \ac{SOC} estimation, the energy buffers introduced in \eqref{eq:fcr_s_max} and \eqref{eq:fcr_s_min} guarantee that a worst-case \ac{FCR} activation can always be delivered. Under a \ac{SOC} error $w_t$, however, the energy margins need to be adjusted, namely:
\begin{align}
    \tilde{s}_{\min,t} \le s_t^{\text{rep}} + w_t \le \tilde{s}_{\max,t}. \label{eq:ideal_max}
\end{align}
Since $w_t$ is unobservable, we employ a constraint-tightening approach and define an additional margin $m_t$ to hedge against the uncertainty on the \ac{SOC}:
\begin{equation} \label{eq:final_bidding_problem}
\begin{aligned}
\max \quad & J \\
\text{s.t.} \quad & s_{t}, s_{t+1} \leq \tilde{s}_{\max,t} - m_t, \\
& s_{t}, s_{t+1} \geq \tilde{s}_{\min,t} + m_t, \\
& \text{Battery and market constraints.}
\end{aligned}
\end{equation}
The constraint-tightening parameter $m_t$ is selected such that the true \ac{SOC} stays within the limits of $[\tilde{s}_{\min,t}, \tilde{s}_{\max,t}]$ in closed-loop operation for all times $t$ over a horizon $N$ with a joint probability of at least $1-\alpha$:
\begin{equation}\label{eq:chance_constraint1}
    \mathbb{P} \left( \bigcap_{t=1}^N \left\{ \tilde{s}_{\min,t} \le s_t^{\text{rep}} + w_t \le \tilde{s}_{\max,t} \right\} \right) \ge 1 - \alpha.
\end{equation}
For $\alpha = 0$, this corresponds to a robust constraint-tightening. Note that $N$ can differ from the prediction horizon $T$ and serves as a tuning parameter for the desired robustness of \eqref{eq:chance_constraint1}. Under the approximation that the optimization variables $s_t$ and the reported \ac{SOC} $s_t^{\text{rep}}$ are equal, the optimizer consequently restricts the reported state $s_t^{\text{rep}}$ to a tightened feasible region $[\tilde{s}_{\min,t} + m_t, \tilde{s}_{\max,t} - m_t]$. Hence, \eqref{eq:chance_constraint1} is satisfied if the error is bounded by the margin $m_t$ over the horizon $N$ with a probability of $1-\alpha$:
\begin{equation}
    \mathbb{P} \left( \bigcap_{t=1}^N \left\{ |w_t| \le m_t \right\} \right) \ge 1 - \alpha. \label{eq:chance}
\end{equation}
An analytical solution to this chance constraint is hindered firstly by the nonlinearity and nonsmoothness of the error dynamics. Second, the error process is decision-dependent, i.e., there is a coupling between the optimizer's decisions and the evolution of the error. Consequently, we cannot analytically solve~\eqref{eq:chance} for $m_t$ or ensure compliance with a specific value for $\alpha$. Rather, we tune the tightening in simulation over a chosen horizon $N$ to achieve sufficiently robust performance. In the following, we present three different approaches of increasing complexity to choose the margin $m_t$.

\subsubsection{Naive, fixed-margin optimization}
The simplest approach to robustly account for the uncertainty $w_t$ is to choose a constant constraint tightening:
\begin{equation}
    m_t = m.
\end{equation}
The larger we choose the constant $m$, the higher the error that can be absorbed. In fact, choosing $m \geq w_{\max}$ guarantees robust constraint satisfaction. However, introducing such a fixed constraint tightening does not account for the varying nature of SOC uncertainty and leads to conservative bidding and operation of the battery.

\subsubsection{Adaptive-margin optimization}
Instead of a fixed margin, this approach uses a time-dependent tightening that is adjusted based on the system's realized trajectory according to:
\begin{equation}
    m = m(t) = 
    \begin{cases}
        m(t-1) + \Bar{w}, & \text{if } b < s(t-1) < c,\\[4pt]
    \gamma m(t-1) & \text{otherwise}.
    \end{cases}
\end{equation}
This accounts for the fact that estimation certainty is regained near the SOC boundaries, as operating at high or low SOC levels allows for a more accurate estimation:
The margin $m(t)$ is assumed constant for the horizon of the optimization, but is recalibrated based on the realized trajectories before each optimization step. The constant $\gamma \in [0,1]$ governs the decrease in margin, whereas $\Bar{w}$ represents the increase in the margin at each step. These need to be calibrated such that $m(t)$ robustly encapsulates $w_t$.

\subsubsection{Uncertainty-aware optimization}
This approach treats the tightening margin as dependent on the dispatch decisions. Drawing from homothetic tube MPC \cite{Rako2012}, we introduce an uncertainty set $[-\delta_t, \delta_t]$, whose predicted scaling $\delta_t$ is a decision variable:
\begin{equation}
    \begin{aligned}
        \min\quad & J\\
\text{s.t.} \quad & s_{t}, s_{t+1} \le \tilde{s}_{\max,t} - \delta_t, \label{eq:final_s_max} \\
& s_{t}, s_{t+1} \ge \tilde{s}_{\min,t} + \delta_t, \\
& \delta_0 = \delta (t),\\
& \delta_{t+1} = 
\begin{cases}
\delta_t + \Bar{w}, & \text{if } b < s_t < c,\\[4pt]
\gamma \textcolor{black}{\delta_t} & \text{otherwise},
\end{cases} \\
& \text{Battery and Market Constraints}.
    \end{aligned}
\end{equation}
This uncertainty-aware formulation allows the optimizer to take decisions to actively reduce the constraint tightening $\delta_t$. Deliberately steering the SOC trajectory toward high or low regions reduces uncertainty and narrows the margin. By choosing the increase in margin $\Bar{w}$ large enough, the tube $\Delta = [-\delta_t, \delta_t]$ eventually encapsulates the error, i.e., the chance-constraint \eqref{eq:chance} is satisfied for a sufficiently small $\alpha$.

\section{Case Study: German \ac{DAM} and \ac{FCR} Market}\label{sec:caseStudy}
We evaluate the three different constraint-tightening approaches in a case study on the German \ac{DAM}. We assume a \ac{BESS} with a rated power of $P_{\max} = 10$ MW and installed energy capacity of $C = 10$ MWh, corresponding to a 1-hour battery system. We simulate the simultaneous bidding on the \ac{DAM} and the \ac{FCR} market with an hourly resolution over a 6-months period in 2024. The spot market prices correspond to the historic \ac{DAM} prices in Germany, whereas for simplicity, we assumed a constant price for \ac{FCR} provision. An overview of the chosen parameters is given in Table \ref{tab:parameters}.

\begin{table}[t]
\centering
\caption{\textsc{Case study parameters}}
\label{tab:parameters}
\begin{tabular}{lll}
\hline
\textbf{Parameter} & \textbf{Symbol} & \textbf{Value} \\ \hline
Rated Power, Nominal Capacity & $P_{\max}$, $C$ & 10 MW, 10 MWh \\
Charging/Discharging Efficiency & $\eta_{\text{ch}}$, $\eta_{\text{dis}}$ & 0.99, 0.99 \\
Time Step, Prediction Horizon & $\Delta t$, $T$ & 1 hour, 72 hours \\
SOC Error Parameters & $b$, $c$ & 0.15, 0.9 \\
SOC Error Uncertainty & $\beta$, $\sigma^2$ & 1e-3, 1e-4\\
Penalty Degradation & $c_{\text{deg}}$ & 36.5\euro/MWh\\
Penalty Imbalance & $c_{\text{imb}}$ & 1e4\euro/MWh\\
Penalty FCR Shortfall & $c_{\text{fcr}}$ & 1e5\euro/MWh\\
FCR Price & $\pi^{\text{FCR}}$ & 16 \euro/MW/h\\ 
SOC Error Decay Rate & $\gamma$ & 0.8\\
Sustained \ac{FCR} Activation & $\Delta t_{\text{FCR}}$ & 30 min \\
\hline
\vspace{-0.6cm}
\end{tabular}
\end{table}

Note that the penalties for degradation, imbalance, and \ac{FCR} shortfall are parameters in the optimization and do not exactly reflect the reality of the incurred costs. They are chosen to ensure numerical stability and a penalty-based hierarchy in the optimization. In contrast, for the evaluation of the bidding strategy, metrics are based on the real economic revenues and costs. The total revenue $R$ is the sum of the revenue from the \ac{DAM} and the \ac{FCR} market, $R_{\text{DAM}}$ and $R_{\text{FCR}}$, from the imbalance settlement $C_{\text{imb}}$ and the degradation cost $C_{\text{deg}}$, i.e.,
\begin{equation}
    \begin{aligned}
        R &= R_{\text{DAM}} + R_{\text{FCR}} + C_{\text{imb}} - C_{\text{deg}}, \\
        R_{\text{DAM}} &= \Delta t \sum_{t=1}^T \pi_t^{\text{DA}} \cdot p^{\text{DA}}_{\text{bid},t}, \\
        R_{\text{FCR}} &= \Delta t \sum_{t=1}^T \pi_t^{\text{FCR}} \cdot p^{\text{FCR}}_{\text{bid,},t}, \\
        C_{\text{imb}} &= \Delta t \sum_{t=1}^T \pi_t^{\text{IB}} \cdot p_{\text{imb, t}}^{\text{true}}, \\
        C_{\text{deg}} &= \Delta t \sum_{t=1}^T c_{\text{deg}} \cdot p_{\text{dis,t}}^{\text{true}}. 
    \end{aligned}
\end{equation}

The imbalance settlement follows a dual pricing scheme \cite{Vande2010}, where the imbalance price is 30\% higher or lower than the spot market price, depending on whether the asset needs to buy or sell on the balancing market, respectively. This ensures that the deviations from the scheduled power dispatch remain economically unfavorable. Furthermore, we quantify battery degradation using an equivalent-cycle model with the same degradation cost $c_{\text{deg}}$ as used in the objective function in the optimizer \cite{Xu2022}.

In addition, we introduce the metrics of \ac{FCR} compliance and FCR energy shortfall. The first reflects whether during a certain time interval, the \ac{BESS} is capable of sustaining a worst-case FCR activation given the true SOC and the amount of reserves it has promised. The latter quantifies, in the event of noncompliance, how large the energy shortfall is. Note that we do not include explicit costs of this in the revenue $R$, since this is a measure that is not typically observed in reality, where full activations of \ac{FCR} are rare.

\section{Results}\label{sec:results}
In this section, we evaluate and discuss the performance of the different constraint-tightening strategies in the presented case study.

\subsection{Analysis of Error Dynamics and Constraint Tightenings}
\begin{figure}[t]
    \centering
    \includegraphics[width=0.9\linewidth]{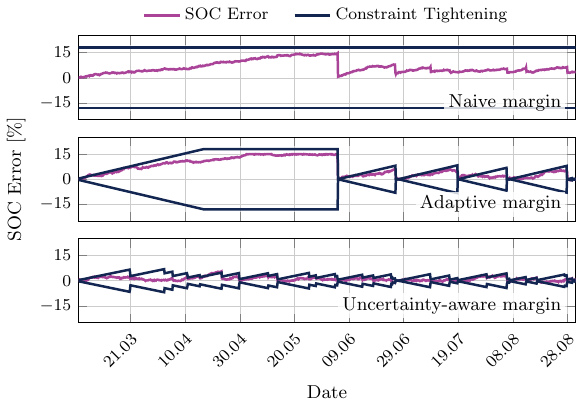}
    \caption{SOC error and constraint tightening for the three different approaches.}
    \vspace{-0.5cm}
    \label{fig:constrainttightening}
\end{figure}

First, we analyze the dynamics of the error and the effectiveness of the chosen constraint tightening. Figure \ref{fig:constrainttightening} shows the evolution of the \ac{SOC} error and the constraint-tightening tube $[-m_t, m_t]$ that result from the market simulation using the three different approaches.

Under both the naive and adaptive approaches, the battery initially operates continuously in the \ac{FCR} market, maintaining its \ac{SOC} around 50\%. During this period, the estimation error accumulates and begins to drift. The \ac{SOC} error decreases whenever it is economically profitable to participate in the \ac{DAM} and fully charge or discharge the battery. 

While a sufficiently large fixed margin $m$ ensures that the error is encapsulated robustly in the naive approach, it introduces excessive conservatism, and effectively "blocks" a significant portion of the \ac{SOC} range. In contrast, the adaptive-margin approach results in a tighter tube by contracting the margin whenever the optimizer has deemed a \ac{DAM} cycle economically profitable. However, this approach remains reactive, i.e., it does not actively reduce the uncertainty, but rather allows the margin to grow to high values. Consequently, a large portion of SOC, which could otherwise be used for market participation, is again reserved to hedge against the uncertainty of the SOC.

In contrast, the uncertainty-aware optimizer explicitly chooses to participate in the \ac{DAM} more often to reduce the uncertainty and, therefore, the margin in \ac{SOC}. This leads to a less conservative margin on average, which, however, still robustly captures the error with high probability. \added{Furthermore, the approach is computationally efficient, with a single operational planning step solving in an average of 0.08 seconds on a standard notebook (Intel Core Ultra 7 258V CPU, 32~GB RAM).}

\subsection{Revenue and Performance Analysis}
The different levels of conservativeness can also be observed in the total revenue. Table~\ref{tab:algorithm_performance} compares the different metrics between an optimizer without constraint tightening and the three proposed approaches, all tuned to achieve \ac{FCR} compliance in more than 96\% of the hours.

\begin{table*}[h] 
\centering
\vspace{-0.2cm}
\caption{\textsc{Performance Comparison}}\label{tab:algorithm_performance}
\begin{tabularx}{\textwidth}{@{} l XXXX @{}} 
\toprule
\textbf{Metric} & \textbf{No Tightening} & \textbf{Fixed-Margin Approach} & \textbf{Adaptive-Margin Approach} & \textbf{Uncertainty-Aware Approach} \\
\midrule
\quad \textbf{Total Net Revenue (k\euro)}        & 842.7 & 573.0 & 648.4 & 786.0 \\
\midrule
\quad \textbf{FCR Revenue (k\euro)}                  & 782.2  & 530.6  & 606.8  & 716.6  \\
\quad \textbf{DAM Revenue (k\euro)}               & 70.0  & 49.0  & 45.9  & 82.5  \\
\quad \textbf{Imbalance Settlement (k\euro)}         & -2.1    & -1.1    & -0.7    & -0.9   \\
\quad \textbf{Degradation Cost (k\euro)}             & -7.4  & -5.5  & -3.7  & -12.1  \\
\midrule
\quad \textbf{Cumulated Shortfall Energy (MWh)}       & 1267.9      & 0.0      & 16.4      & 4.3     \\
\quad \textbf{Shortfall Hours (\%)}          & 92.7      & 0.0      & 3.8       & 1.9      \\
\midrule
\quad \textbf{Mean Margin}                  & 0.0      & 0.160      & 0.116      & 0.029      \\
\quad \textbf{Parameter}            & $m$ = 0.0 & $m$ = 0.160  & $\bar{w}$ = 0.00018 & $\bar{w}$ = 0.00018\\
\bottomrule
\vspace{-0.41cm}
\end{tabularx}
\end{table*}

Clearly, not including any margin to account for \ac{SOC} error results in the highest \ac{SOC} utilization for bidding. Consequently, revenues from the different markets are highest. However, not introducing any margin results in non-compliance with the submitted bids in over 90\% of the hours. Note that even a small violation is considered non-compliant here. While not directly reflected in the total net revenue, this is clearly an operational risk, both from a system perspective and from an asset owner perspective, who runs the risk of losing prequalification due to noncompliance with promised reserves. Therefore, it is essential to introduce margins to account for \ac{SOC} uncertainty in battery bidding.

The fixed-margin approach with sufficiently high $m$ robustly reduces the number of shortfall hours to zero. However, its conservativeness directly translates into reduced total revenue. The adaptive-margin approach is less conservative, with a lower mean margin, while still maintaining compliance with promised reserves at almost all times. Not only is the number of shortfall hours drastically reduced compared to the case with no tightening, but also the cumulated energy shortfall is negligibly small.

The uncertainty-aware approach, on the other hand, performs best in terms of the trade-off between total revenue and FCR compliance. Its revenue is closest to the untightened case, while exhibiting only a small FCR shortfall in energy and hours. It achieves this by leveraging the decision-dependent nature of SOC uncertainty. It accepts suboptimal DAM dispatch and reduced FCR participation in certain intervals in order to reduce the margin in favor of future profits. On the one hand, this is reflected by an increased \ac{DAM} revenue. On the other hand, it also trades off a higher degradation and less participation in the FCR markets in favor of the lower mean margin.

\subsection{Levels of Robustness and FCR Compliance}
The uncertainty-aware constraint tightening demonstrates a favorable trade-off between revenues and \ac{FCR} compliance. To illustrate that this is a merit of the uncertainty-aware formulation, we repeat the analysis of the previous section for various tunings of the different approaches. More precisely, we simulate the hourly participation of the fixed-margin strategy with varying tightening parameters $m\in [0,0.18]$ in equidistant steps of 0.03. Similarly, we simulate the adaptive-margin and uncertainty-aware approaches for parameter values $\bar{w} \in [8e^{-5}, 18e^{-4}]$ in equidistant steps of $2e^{-5}$. \added{These choices correspond to a total growth of approximately 4\% to 9\% in the absolute SOC margin over a 3-week horizon.}

\begin{figure}
    \centering
    \includegraphics[width=0.85\linewidth]{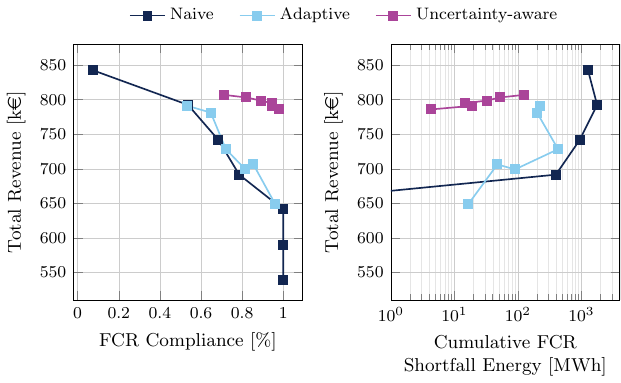}
    \caption{Left: Trade-off between total revenue and FCR compliance. Right: Trade-off between the total revenue and the cumulative FCR shortfall energy.}
    \vspace{-0.5cm}
    \label{fig:paretoboth}
\end{figure}
The resulting trade-offs between total revenue and FCR compliance and cumulative FCR shortfall energy are shown in Fig.~\ref{fig:paretoboth}. The fixed-margin and adaptive-margin optimizers perform approximately the same in terms of number of compliant hours, while the adaptive-margin manages to reduce the amount of FCR shortfall energy, i.e., the magnitude of constraint violation. However, it is obvious that they both perform worse than the uncertainty-aware optimizer over the whole region of high FCR compliance. At a constant 3\% tightening, for instance, the fixed-margin approach does not reliably capture the \ac{SOC} uncertainty and is compliant in only about half of the hours. In contrast, at a similar revenue, the uncertainty-aware formulation shows compliance in more than 98\% of hours. This clearly showcases the significance of incorporating the decision-dependent nature of the \ac{SOC} uncertainty into the operational strategy.

\section{Conclusion}\label{sec:conclusion}
In this paper, we study how \ac{SOC} uncertainty affects the market participation of \ac{LFP} \acp{BESS}. We demonstrate that ignoring \ac{SOC} uncertainty in the bidding strategy leads to increased non-compliance with submitted bids and an increased risk of failing to sustain a worst-case activation of \ac{FCR}. We propose three constraint-tightening reformulations of the bidding problem that reduce the number of such non-compliant hours: a constant constraint tightening, an adaptive constraint tightening, and an uncertainty-aware constraint tightening on the \ac{SOC}. The latter allows to strategically leverage the decision-dependent nature of SOC uncertainty and to plan actions to reduce it. Although all three approaches can be used effectively to robustify the bidding against SOC errors, the uncertainty-aware formulation yields higher revenues at the same compliance level than the constant or adaptive approaches. This demonstrates the benefit of treating SOC uncertainty as an endogenous process within the operational strategy.

\added{While the numerical results are specific to the evaluated test case, the underlying methodology of treating uncertainty as decision-dependent remains generalizable across different market settings and battery applications. Future research will explore these broader environments and conduct detailed sensitivity analyses on the error model and parameters. Furthermore, since the constraint-tightening margins are currently tuned heuristically, future work will focus on deriving probabilistic guarantees for constraint satisfaction and quantifying the associated operational risk.}

\section*{Acknowledgments}
{\footnotesize This work was supported by the NCCR Automation, a National Centre of Competence in Research, funded by the Swiss National Science Foundation (grant number 51NF40\_225155). The authors also wish to thank Dr. Georgios Darivianakis for his valuable support and insights throughout this project.\par}

\bibliographystyle{IEEEtran}


\end{document}